# The Machine Can Say It but Cannot Hear It: Designed Affective Patterns and the Expressive–Sensing Asymmetry in Human–Machine Communication

Jan K. Argasiński
*Institute of Applied Computer Science,*
*Faculty of Physics, Astronomy and Applied Computer Science,*
*Jagiellonian University in Krakow, Poland*
*& Nano Games sp. z o.o., Krakow, Poland*
*jan.argasinski@uj.edu.pl*

## Abstract

Affect-adaptive systems increasingly act as communicators that sense a user's emotion and respond with events meant to change it, closing an affective loop. This vision assumes both that a machine's affective messages are received and that the bodily channel it monitors carries an intelligible reply—assumptions rarely tested together. In a within-subjects virtual-reality study (N = 20), an autonomous system delivered six empirically derived affective patterns—scripted emotional events distilled from 104 practitioners' (first responders') critical incidents—while we recorded the human reply across felt emotion, felt arousal, and autonomic (electrodermal and cardiac) activity. Acting only as an author of designed messages, the machine reliably evoked strong, differentiated emotions: valence fell sharply for every pattern ($|d_z|$ = 1.1–1.7), and the patterns produced distinguishable, individually classifiable signatures of anger, fear, and sadness, functioning as a vocabulary of machine-to-human affective messages. Yet the channel the machine would read stayed largely silent. Self-reported arousal did not change, with Bayesian and equivalence evidence for no effect; sympathetic and cardiac arousal moved only for the most perceptually salient events, not for the messages experienced as most powerful; and within individuals, felt valence was decoupled from both bodily registers. This valence–arousal dissociation reveals an expressive–sensing asymmetry: machine agency extends to expression but not perception, and closed-loop designs regulate on a variable their messages neither reliably move nor are legible in. We draw out implications for theories of machine agency and for affect-adaptive design.

# Introduction

A growing class of communication technologies no longer merely transmits messages between people; it composes messages of its own and adapts to the people who receive them. Conversational agents, social robots, recommender systems, and affect-adaptive interfaces behave less like channels and more like interlocutors, initiating actions to which users respond socially and emotionally. This shift is the animating premise of the human–machine communication (HMC) research program and of a broader reorientation within computer-mediated communication scholarship from studying communication *through* computer-based media to studying communication *with* them (Guzman & Lewis, 2020; Sundar, 2020). When the machine is treated as a communicative source rather than a conduit, familiar questions about message design, uptake, and feedback can be asked of the machine's own messages—including, increasingly, its emotional ones.

Among the most ambitious systems in this class are those that attempt to *close an affective loop* with the user. Such systems continuously infer a person's emotional state from behavioral or bodily signals and, in response, deliver events intended to move that state toward a desired range—maintaining engagement in a tutor, calming a user of a therapeutic application, sustaining challenge in a game, or preserving realistic stress in a training simulation (Picard, 1997). Described in engineering terms this is a control loop; described in communicative terms it is a dialogue. The machine "listens" to the body, "speaks" by staging an affective event, and "listens" again for the reply. The premise that makes the loop coherent is fundamentally communicative: the machine's affective messages are received and the channel it monitors carries an intelligible answer.

That premise is rarely examined directly, and almost never on both sides at once. A large literature shows that people respond to machine-authored cues as if they were social and emotional acts (Nass & Moon, 2000; Reeves & Nass, 1996), and a separate literature in affective computing catalogs the bodily correlates of emotion (Kreibig, 2010). But whether a machine's *designed* affective messages reliably change a user's felt emotion, and whether the autonomic signals the machine reads back actually track that change, are distinct empirical questions that must be tested together and multimodally to evaluate the loop as a communicative system. If the two sides come apart—if the machine can send affect but cannot sense the reply it elicits—then a class of systems now being deployed rests on an unwarranted assumption.

The present study examines both sides in a single, reproducible design. A machine delivers a battery of empirically derived *affective patterns*—scripted, emotionally charged events—to participants immersed in a virtual-reality scenario, and we measure the human reply on three registers: the felt *quality* of emotion (discrete emotions and affective valence), felt *arousal*, and *autonomic* arousal recorded from low-cost electrodermal and cardiac sensors. The first finding is affirmative and, we argue, of independent interest: acting purely as an author of designed messages, the machine reliably evokes strong and *differentiated* emotions—anger to an aggressive bystander, fear to an explosion, sadness and guilt to a miscarriage—so that the patterns function as a small vocabulary of machine-to-human affective messages rather than as a single undifferentiated stressor. The second finding is a pointed asymmetry. The very same messages that reshape felt emotional quality leave felt arousal unchanged and move autonomic arousal only for a minority of salient, startle-like events—crucially, not for the messages that participants experienced as most emotionally powerful. The register the machine moves is thus decoupled from, and far larger than, the registers it would read.

This paper makes three contributions to the study of affective human–machine communication. First, it reframes *affective patterns* as a reusable, ecologically grounded vocabulary for machine-initiated affective messaging and validates that vocabulary multimodally, showing that an autonomous system can author differentiated emotional meaning, not merely generic activation. Second, it documents a valence–arousal *dissociation* supported by an informative null: using Bayesian and equivalence methods, we provide positive evidence that the machine's messages do *not* change felt arousal, rather than merely failing to detect a change. Third, it draws from this dissociation a generalizable constraint on closed-loop affective communication—an *expressive–sensing asymmetry* that holds across self-report, sympathetic (electrodermal), and parasympathetic (cardiac) channels—and derives design implications for systems that would regulate communication on a bodily signal. The contribution is theoretical and methodological rather than a demonstration of any particular application.

The remainder of the paper proceeds as follows. We first situate the affective loop within the turn toward communicating *with* machines and the psychology of human–AI interaction, motivating the expectation that machine-authored affective messages are genuinely received (H1–H2). We then draw on dimensional and discrete models of emotion, and on the problem of autonomic specificity, to motivate the expectation of a channel mismatch on the sensing side (H3–H4, RQ1–RQ2). After describing a within-subjects virtual-reality study, we report the dissociation and its convergence across registers, and discuss what an expressive channel that works and a sensing channel that does not imply for the design of affect-adaptive communication.

## Theoretical Background and Hypotheses

### *From Communicating Through to Communicating with Machines*

Computer-mediated communication has historically treated the computer as a medium situated between human interlocutors. The rise of autonomous and semi-autonomous systems has unsettled that arrangement: when software authors its own messages and adapts to a user's responses, the machine occupies the role of a communicative participant, and the appropriate object of study becomes the interaction between a person and the system itself (Guzman & Lewis, 2020). This reframing invites us to treat a system's affective behavior as *message design*—a set of choices about what to express, how, and when—and to treat the user's bodily and self-reported responses as the *reply* in an exchange. An affect-adaptive system that senses a user and stages an emotional event is, on this view, engaged in a bidirectional affective exchange: an expressive act directed from machine to human, and a sensing act by which the human's state is read back. Evaluating such a system as communication requires asking, of each direction, whether the message lands and whether the reply is legible.

### *Machine Agency and Social Responses to Machine-Authored Acts*

Two complementary traditions predict that machine-authored affective messages will be genuinely received. The Computers Are Social Actors paradigm and the media equation demonstrated that people apply social rules and emotional responses to computers that exhibit even minimal social cues, responding to scripted machine behaviors "mindlessly" as though to a social other (Nass & Moon, 2000; Reeves & Nass, 1996). More recently, Sundar's (2020) framework for the psychology of human–AI interaction locates the source of these effects in perceived *machine agency*: as systems are seen to act on their own, users attribute intention and source-hood to them, with downstream consequences for how the systems' outputs are experienced. If a machine is treated as an agentic source, then the affective events it authors should be experienced as emotional communications and should produce real emotional uptake in the receiver. We therefore expect designed affective patterns to shift users' felt emotion relative to a baseline (H1). We use *agency* here in the interactional sense employed by the social-actor tradition—the system as the proximate source of the messages user responds to—rather than as a claim about autonomous real-time generation; the patterns in this study are pre-authored, exactly as the scripted stimuli in classic media-equation experiments were, and are delivered by the system as events within the interaction.

### *Affective Patterns as a Message-Design Vocabulary*

If machine-authored affect is received, a natural next question is whether it can be *designed* with specificity—whether a system can compose distinguishable emotional messages rather than merely raising or lowering generic activation. The construct of the *affective pattern* was introduced to serve this purpose: a reusable unit that couples a designed event to an intended affective response, merging traditions of game-design patterns with evidence-centered approaches to modeling the user (Argasiński & Węgrzyn, 2019). A vocabulary is only useful if its elements are discriminable, and the ecological provenance of the patterns studied here bears directly on that requirement. They were distilled from a qualitative study of 104 emergency-service practitioners who described 919 critical incidents, condensed into twenty modeled events and, in the present study, six selected for eliciting distinct emotional profiles. This grounding in real, emotionally consequential situations gives the patterns a claim to ecological validity uncommon in laboratory affective stimuli, and motivates the hypothesis that they evoke *differentiated* discrete emotions—recognizable, pattern-specific signatures—rather than a uniform negativity (H2). Differentiation is what would license treating affective patterns as a communicative vocabulary that could be reused across settings such as tutoring, therapy, and games, and not merely as a domain-specific stressor.

*Dimensional and Discrete Emotion, and the Problem of the Sensed Channel*

The two directions of the affective loop are typically indexed by different features of emotion, and this is where a mismatch may arise. Dimensional models describe affect along the axes of valence (pleasantness) and arousal (activation), operationalized in self-report by instruments such as the Self-Assessment Manikin (Bradley & Lang, 1994; Russell, 1980). Closed-loop affective systems, however, almost always regulate on *arousal*, because arousal is taken to index engagement and to have a legible autonomic correlate: electrodermal activity as a marker of sympathetic activation, and heart-rate variability as a marker of parasympathetic activation (Kreibig, 2010; Picard, 1997). The design of such systems therefore stakes its sensing side on the assumption that the machine's affective messages move arousal and that arousal is faithfully expressed in the autonomic signal.

Both links in that assumption are contestable. First, designed narrative events carry emotional *meaning*—loss, threat, injustice—that may register most strongly on the valence and discrete-emotion axes while leaving the arousal axis comparatively untouched; a message can be devastating without being activating. Second, and more fundamentally, the mapping from emotion to autonomic state is many-to-one: decades of psychophysiology have failed to establish consistent autonomic "fingerprints" for discrete emotions, and meta-analytic evidence indicates that peripheral signals underdetermine emotional state (Cacioppo et al., 2000; Kreibig, 2010; Siegel et al., 2018). If the messages move quality more than arousal, and if arousal is in any case an unreliable readout of felt emotion, then a system that listens on the autonomic channel may be deaf to precisely the emotional communication it succeeds in sending. We therefore test whether the patterns shift self-reported arousal (H3) and whether they shift autonomic arousal—electrodermal and cardiac—against a fair, activity-matched baseline (H4), and we ask how the expressive and sensed registers co-vary within individuals (RQ1) and whether low-cost cardiac measurement can serve as a usable parasympathetic channel at all (RQ2). Consistent with the dimensional logic above and the specificity problem, we anticipate that H3 will yield an *informative* null and that H4 will be supported, if anywhere, only for the most perceptually salient events.

### *Psychological Fidelity: The Use Case, Not the Claim*

The scenario used here—virtual-reality rescue training—supplies ecologically strong, negatively valenced stimuli for a principled reason, which we state briefly because it motivates the materials without constituting the paper's contribution. Training simulations aim not only at physical fidelity but at *psychological fidelity*: evoking the emotional responses of the real task so that trainees rehearse coping and avoid maladaptive reactions under stress (Hontvedt & Øvergård, 2020; Wild et al., 2020). This rationale justifies designing patterns that are genuinely affecting and drawing them from practitioners' most consequential experiences. It does not bear on the communicative questions at the center of this paper, and we make no claim about training efficacy; the value of the scenario for present purposes is that it provides emotionally potent, real-world-grounded messages with which to test the two sides of the affective loop.

### *Summary of Hypotheses and Research Questions*

- **H1 (expressive channel).** Machine-delivered affective patterns shift participants' self-reported emotional quality and valence relative to baseline.
- **H2 (message vocabulary).** The patterns evoke *differentiated*, pattern-specific discrete-emotion profiles rather than a single undifferentiated negative shift.
- **H3 (sensed channel, self-report).** The patterns shift self-reported *arousal*. Given the dimensional logic and the specificity problem, we test H3 as a directional expectation while anticipating, and evaluating with Bayesian and equivalence methods, an *informative null*.
- **H4 (sensed channel, physiology).** The patterns shift autonomic arousal—electrodermal (sympathetic) and cardiac (parasympathetic)—against an activity-matched baseline; we anticipate support, if any, only for the most perceptually salient events.
- **RQ1 (convergence).** Within individuals and across patterns, how do felt valence, felt arousal, and autonomic arousal co-vary?
- **RQ2 (feasibility).** Can low-cost, short-segment cardiac measurement serve as a dependable parasympathetic "listening" channel in this setting?

# Method

## *Design*

We used a within-subjects design with seven repeated affective states: a baseline measurement and one measurement after each of six machine-delivered affective patterns. The patterns were embedded, in an order randomized across participants, within a single continuous virtual-reality mass-casualty scenario (a bus crash on an airport apron requiring triage of multiple victims). Because every participant encountered every pattern, each person served as their own control, yielding six within-person pattern-versus-baseline contrasts per dependent measure.

## *Participants*

Participants were 20 general-population adult volunteers (12 men, 8 women; $M_{\text{age}} = 25.6$, $SD = 3.8$, range 20–34). Most had little or no prior virtual-reality experience (14 had used VR only in isolated sessions, 5 had none, and 1 reported a single prior exposure). None were emergency-services professionals; as detailed below, the affective patterns were derived from a separate practitioner sample. Reported spatial presence was high (Igroup Presence Questionnaire spatial-presence subscale $M = 5.56$), indicating that participants experienced the scenario as immersive.

## *Materials: The Affective-Pattern Library*

The stimuli were *affective patterns*—scripted, emotionally charged events staged within the scenario. Their derivation was ecological rather than intuitive. In a preceding qualitative study, 104 paramedics and firefighters described the most stressful events they had encountered on operations, yielding 919 critical-incident samples that were distilled into 20 modeled events and implemented in the simulation. Two pilot studies (34 practitioners in total) screened these events, and six were retained for the present study on two criteria: they produced the strongest physiological reactions in piloting, and they were expected, from their content, to elicit *qualitatively distinct* emotions. The six retained patterns were an aggressive bystander, an explosion, a screaming and panicking victim, bystanders recording the scene on phones, a pregnant victim suffering a miscarriage, and a scream. Each was realized as an audiovisual event within the ongoing triage task.

### *Apparatus*

The scenario ran on an established disaster-simulation VR engine driving a head-mounted display. Physiology was recorded continuously with a low-cost, open-hardware acquisition device (BITalino), capturing electrodermal activity and single-lead electrocardiography at 100 Hz. The choice of low-cost hardware was deliberate, reflecting the deployability goals of affect-adaptive systems. A separate device-comparison study ($n \approx 20$) indicated moderate, participant-dependent agreement between this hardware and a research-grade reference; we therefore treat the acquisition as *adequate for within-subject contrasts*, not as metrologically validated, and interpret the physiological results accordingly.

### *Measures*

After the baseline and after each pattern, participants reported their affect on two instruments: the Self-Assessment Manikin (Bradley & Lang, 1994), providing 9-point ratings of valence, arousal, and dominance; and a six-emotion mood scale (Wojciszke & Baryła, 2005), providing 7-point ratings of anger, fear, guilt, joy, love, and sadness. Continuous electrodermal activity and electrocardiography were segmented into the pattern events and their matched baselines (see Analysis). Simulator sickness was assessed once with the Simulator Sickness Questionnaire (Kennedy et al., 1993) as a control for discomfort-based confounds, and spatial presence with the Igroup Presence Questionnaire.

### *Analysis*

Each pattern was compared with baseline using paired $t$ tests, with Cohen's $d_z$ as the standardized effect size. To evaluate evidence *for* as well as *against* effects—essential for interpreting the arousal channel—we computed Jeffreys–Zellner–Siow Bayes factors for each contrast (Cauchy prior scale $r = .707$), reporting $BF_{10}$ for expected effects and $BF_{01} = 1/BF_{10}$ for null-direction tests, with robustness checks across prior widths ($r = .5$–$1.0$). For the arousal contrasts we additionally performed two one-sided equivalence tests (TOST); because no external smallest-effect-of-interest was available, we set the equivalence bound at the effect size the design could detect with 80% power ($d_z = 0.66$), following recommended practice for justifying bounds by sensitivity (Lakens, 2017). Electrodermal responses were quantified against three baselines— passive rest, the pre-trial active state, and a per-pattern activity-matched baseline—with the *matched-active baseline pre-designated as the fair test*; results were verified across three preprocessing pipelines. Heart-rate variability was recomputed from the raw electrocardiogram with a documented quality-control pipeline (bandpass filtering, R-peak detection, ectopic-beat correction, and a signal-quality gate), reporting the usable-segment fraction and time-domain indices. Within each dependent-variable family of six pattern contrasts, $p$ values were corrected for multiple comparisons using Holm's procedure (family-wise) and Benjamini–Hochberg (false-discovery-rate), reported alongside the Bayes factors. Within-person convergence among registers was assessed with repeated-measures correlation (Bakdash & Marusich, 2017). Exact per-test $n$ is reported throughout, because signal-quality exclusions reduced the electrodermal sample for some patterns.

# Results

## *The Machine Authors Differentiated Emotion*

The expressive channel worked with specificity. Every pattern produced a large, highly reliable drop in affective valence relative to baseline (all $p < .001$; $|d_z| = 1.11–1.72$; $BF_{10}$ from $3.1 \times 10^2$ to $5.6 \times 10^4$), together with decreases in joy and increases in anger, fear, and sadness (Table 1). All valence effects survived both Holm and Benjamini–Hochberg correction. The most potent message was the miscarriage (valence $d_z = -1.72$).

Crucially, these were not interchangeable negativity inductions. A pattern (6) × emotion (6) repeated-measures analysis of the discrete-emotion change scores revealed a strong interaction—the emotion evoked depended on the pattern—that survived a conservative Greenhouse–Geisser correction, $F(25, 475) = 6.65$, $p < .001$ (Greenhouse–Geisser $p = 2.7 \times 10^{-4}$), $\eta^2_G = .08$. Each pattern carried a recognizable signature (Figure 1): the aggressive bystander loaded on anger, the explosion on fear, the miscarriage on sadness and guilt, while being filmed on phones produced anger *without* fear—the only pattern that did not raise fear, consistent with indignation rather than threat. The signatures were separable at the individual level: a leave-one-participant-out nearest-centroid classifier assigned participants' six-emotion responses to the correct pattern 35.0% of the time against a 16.7% chance rate (binomial $p = 9 \times 10^{-7}$). The most distinct pair of messages—being filmed versus the miscarriage—lay far apart in emotion space (Euclidean profile

distance = 3.40), while panic and the scream were nearest neighbors (0.93). Machine-authored affective patterns thus behaved as a differentiated vocabulary of emotional messages (H1, H2 supported).

### *The Arousal Reply Is Absent*

The sensing side told a different story, beginning with self-report. No pattern changed self-reported arousal: all six contrasts were non-significant ($p$ = .22–.91) with negligible effect sizes ($|d_z| \leq 0.28$), and after Holm correction every arousal $p$ equaled 1.00. This was not a mere failure to detect. Bayes factors favored the null for all six patterns ($BF_{01}$ = 2.2–4.3 at $r$ = .707, reaching the "moderate" threshold of 3 for four of six, and remaining above 1 across prior widths of .5–1.0), and equivalence tests placed five of the six effects within the region the study was powered to detect ($d_z = \pm 0.66$; all $p < .05$; panic $p = .053$). We report transparently that the design was *not* powered to establish equivalence within a very narrow band—the raw ±0.5-point bound was not met for any pattern—so we anchor the equivalence claim to the study's sensitivity and rest the null primarily on the Bayes factors and on the stark contrast with valence (Table 2). The identical messages that shifted felt valence by more than one standard deviation left felt arousal statistically indistinguishable from no change (H3: informative null).

### *Autonomic Arousal Tracks Salience, Not Meaning*

The bodily signals the machine would monitor behaved like the arousal self-report, with one revealing exception. Against the fair, activity-matched baseline, electrodermal (sympathetic) activity differentiated only the most perceptually salient events: the explosion robustly ($d_z$ = 1.23, $BF_{10}$ = 259), and, more moderately, panic ($d_z$ = 0.76) and the aggressive bystander ($d_z$ = 0.64); the explosion and panic survived Holm correction and all three survived false-discovery-rate control. The remaining patterns did not move electrodermal activity ($BF_{10} < 1$), and for the miscarriage the evidence favored the null ($BF_{01} \approx 3.1$). Comparisons against a *passive-rest* baseline spuriously inflated these effects—five of six patterns appeared to raise electrodermal activity—illustrating why the matched-active baseline is the appropriate test; the pattern of significance was near-identical across the three preprocessing pipelines. Cardiac indices, recomputed from the raw electrocardiogram, showed the same profile: mean heart rate differentiated only the explosion ($d_z$ = −1.33, $BF_{10} = 2.1 \times 10^3$) and the scream ($d_z$ = −0.93, $BF_{10}$ = 63) after correction, and RMSSD and SDNN only the explosion ($d_z \approx 0.78$). No cardiac index moved for the miscarriage (all $BF_{10} < 1.3$). Neither autonomic branch registered the machine's most emotionally powerful messages (H4 supported only for salience-type events; Table 2).

### *The Expressive and Sensed Registers Are Decoupled*

Bringing the registers together within individuals shows the asymmetry directly (Figure 2; RQ1). Across the six patterns, within-person change in felt valence was not reliably associated with change in felt arousal ($r_{rm} = .15$, $p = .13$) or with change in electrodermal activity ($r_{rm} = .13$, $p = .22$), and the two autonomic-adjacent channels converged only weakly with each other ($r_{rm} = .21$, $p = .055$). The magnitudes were as decoupled as the covariances: averaged across patterns, the machine moved felt valence by $|d_z| = 1.32$, but felt arousal by only 0.14 and electrodermal activity by 0.58. The register the system moved was thus both far larger than, and statistically independent of, the registers it would read.

### *Controls and Feasibility*

Two checks rule out mundane explanations. First, simulator sickness was mild (Simulator Sickness Questionnaire $M = 0.54$ on a 0–3 scale) and unrelated to the arousal response ($r = .18$, $p = .45$), to baseline arousal, to the valence effect ($r = -.22$, $p = .36$), or to the electrodermal response ($r = .04$); the arousal null held in both low- and high-sickness subgroups. The absent arousal reply is therefore not an artifact of discomfort. Second, the parasympathetic channel's weakness was not a matter of poor signal (RQ2). The raw single-lead electrocardiogram was of good quality (median R-peak signal-to-noise ratio ≈ 26 dB), and 99% of segments yielded usable time-domain heart-rate variability after ectopic correction; the corruption in the study's original automated output (17% of segments physiologically impossible) reflected a processing failure rather than a sensing failure, as those segments recomputed to normal values. The genuine limitation is temporal resolution: 100-Hz acquisition quantizes inter-beat intervals to 10 ms, below rates recommended for reliable beat-to-beat parasympathetic indices (Singh et al., 2004). Low-cost cardiac measurement is thus recoverable for coarse use but is not a turnkey real-time parasympathetic channel—an independent reason, beyond the null responses above, to doubt the sensing side of the loop.

# Discussion

A machine, acting only as the author of designed messages, reliably communicated differentiated emotion to people—anger, fear, sadness, guilt—so that its affective patterns functioned as a small but legible vocabulary. Yet the autonomic channel such a machine would monitor to complete the loop did not carry the reply. Self-reported arousal did not move at all, with positive evidence for no change; sympathetic and parasympathetic activity moved only for the most perceptually salient events and not for the messages participants experienced as most powerful; and within individuals the register the machine moved was statistically decoupled from, and far larger than, the registers it would read. The machine could say it but could not hear it. We take up in turn what this expressive–sensing asymmetry means for theories of human–machine communication, why the body behaved as it did, and what follows for the design of affect-adaptive systems.

## *Machine Agency Extends to Expression but Not to Perception*

The expressive result strengthens, and the sensing result bounds, the account of machines as social and emotional actors. That people respond to machine-authored cues as social acts is well established (Nass & Moon, 2000; Reeves & Nass, 1996), and recent work locates such responses in perceived machine agency (Sundar, 2020). Our findings extend that account from the attitudinal and perceptual level, where it has largely been demonstrated, to differentiated *felt emotion*: an agentic system that stages designed events does not merely raise generic activation but authors specific emotional meaning that receivers take up. In communicative terms, machine affect can be treated as message design, and it can be designed with a discriminable vocabulary.

The asymmetry, however, identifies a limit that the machine-agency literature has had little occasion to confront. Agency in *expression* does not entail competence in *perception*. A system may be an effective emotional source while remaining a poor emotional listener, because the two directions of the exchange rest on different mechanisms: expression exploits humans' readiness to find meaning in staged events, whereas perception requires that the human's state be legible in whatever signal the machine happens to monitor. We name the specific failure a *regulated-variable mismatch*: the system is built to regulate on arousal, yet its messages neither reliably change arousal nor are their felt effects expressed in it. A communicative loop is coherent only when the machine possesses a reply channel matched to the messages it sends; treating the machine as an agent on the expressive side does not supply one. This is a boundary condition on closed-loop affective communication that is independent of engineering refinement—no improvement in sensor fidelity repairs a channel that is monitoring the wrong variable.

### *Why the Body Tracks Salience Rather Than Meaning*

The physiological pattern is not anomalous; it is what an informed reading of psychophysiology predicts. Peripheral autonomic signals underdetermine emotional state, and meta-analytic evidence finds no consistent autonomic "fingerprints" that distinguish discrete emotions (Cacioppo et al., 2000; Kreibig, 2010; Siegel et al., 2018). Against that backdrop, the striking result is not that most patterns failed to move the autonomic channels but that one class of pattern did. The signals responded to the explosion—a sudden, high-intensity audiovisual event—and, for heart rate, to the scream, with the cardiac deceleration and electrodermal change characteristic of an orienting and defensive response to abrupt stimulation (Graham & Clifton, 1966). What the autonomic channel registered, in other words, was perceptual *salience*—the physical suddenness of the event—rather than the appraised *meaning* that drove the valence and discrete-emotion changes. The miscarriage, devastating in felt emotion, is quiet; being filmed provokes indignation without a startle. The expressive channel is keyed to meaning and the sensed channel to salience, and the two need not coincide. This reframes a result that might look like weak measurement into a substantive claim about which feature of a message each channel carries.

### *Design Implications for Affect-Adaptive Communication*

Four implications follow for systems that would communicate affectively with users. First, *regulate on the axis you can move and measure.* If a system's messages reshape valence, appraisal, and discrete emotion, then its control signal should be drawn from those registers—brief self-report probes, behavioral and linguistic indicators, or appraisal-relevant events—rather than from arousal by default. Designing the loop around arousal because arousal has a convenient sensor inverts the proper order of reasoning. Second, *do not trust any single autonomic index.* The two bodily channels converged only weakly with each other and scarcely at all with felt emotion, so a controller that reads electrodermal activity or heart-rate variability alone will act on a noisy and narrow proxy; where physiology is used, multimodal fusion with behavioral and expressive signals is a minimum requirement. Third, *treat low-cost, single-lead cardiac measurement as unfit for turnkey real-time parasympathetic control.* The raw signal here was of good quality, but the intended indices were compromised by a sampling rate below what beat-to-beat metrics require; deployable HRV demands high quality control and adequate temporal resolution, neither of which can be assumed. Fourth, and more constructively, *the machine's expressive competence is itself a resource.* Because designed affective messages land with specificity, systems can communicate affect deliberately even where they cannot sense it—scaffolding, repairing, or intensifying a user's emotional state by message design, with sensing relocated to channels that actually track the intended change.

### *Affective Patterns as a Reusable Communicative Construct*

Beyond the dissociation, the study contributes the affective pattern as a construct for machine-to-human affective communication. The vocabulary demonstrated here is ecologically grounded, distilled from practitioners' most consequential experiences rather than assembled from convenience stimuli; it is differentiated, evoking recognizable and individually classifiable emotional signatures. Although realized in a training scenario, the construct is not bound to training: the same logic—composing designed events that reliably author specified emotions—applies wherever an autonomous system must communicate affect, including tutoring systems that must sustain engagement, therapeutic applications that must modulate distress, games that must pace tension, and companion agents that must respond to a user's emotional turn. Framing such events as message design, rather than as stimuli or as control inputs, connects them to the communicative questions—what to express, to whom, with what uptake—that the study of communicating *with* machines is positioned to ask.

Taken together, the results recommend a modest but consequential shift in how affect-adaptive communication is conceived. The problem such systems face is less a shortage of expressive power than a mismatch between what they can say and what they can hear. The machine in this study was an articulate emotional author and a poor emotional listener, and closing that gap is a problem of choosing the right reply channel, not of speaking more loudly.

### *Limitations*

Several limitations bound the scope of these claims. The sample was modest and comprised general-population young adults. The within-subjects design and the use of Bayesian and equivalence methods make the reported contrasts—including the informative arousal null—interpretable at this sample size, but the study is nonetheless underpowered to establish equivalence within a very narrow band, and effect magnitudes could differ in an different population. Importantly, the central claim concerns the *relation* between message-evoked emotion and its autonomic expression, which there is no reason to expect is cohort-specific.

The materials and setting also limit generalization. The vocabulary was drawn from a single national practitioner sample and delivered within one virtual-reality scenario, so the portability of specific patterns to other populations and domains is an empirical question, even though the underlying construct is general. On measurement, self-reported emotion and arousal were captured after each event rather than continuously, so a brief arousal spike at onset could in principle have resolved before report; this is a genuine caveat for the self-report arousal null, although the continuously recorded autonomic channels converge with it for all but the salient events. The low-cost sensor occasionally saturated (handled by pre-specified exclusions), and cardiac indices are constrained by the 100-Hz acquisition. Finally, the physiological analyses rest on segment means; the interpretation that the body tracks salience rather than meaning would be sharpened by event-locked, fine-grained dynamics. Above all, the study does not test the closed loop. By design we evaluated the loop's two channels but not its operation—the trigger logic is implementable but we do not show if closing the loop maintains arousal. Given the present results, there is a reason to doubt that regulating on arousal would.

### *Conclusion*

Affect-adaptive systems are increasingly cast as communicators that sense a user's emotion and answer it. Examined as communication, the two directions of that exchange came apart. A machine acting only as an author of designed messages reliably evoked differentiated emotion—an expressive vocabulary that landed with specificity—while the autonomic channel it would monitor to hear the reply stayed largely silent, moving only for perceptually salient events and not for the messages that mattered most, and remaining decoupled from felt emotion within individuals. The contribution is this asymmetry, not the scenario that revealed it: machine affective communication can be strong on expression yet thin on perception, and closing the loop is less a matter of speaking more powerfully than of listening on a channel matched to what the messages actually change. The machine can say it; it cannot, yet, hear it.

## Funding

This research was supported by the Polish National Centre for Research and Development under the grant 'Bioadaptable training simulator for critical infrastructure operators – research and preparation for implementation' (POIR.01.01.01-00-1131/17; the Smart Growth Operational Program, submeasure 1.1.1, received by Nano Games sp. z o.o.).

## Data availability

The data underlying this article will be shared on reasonable request to the corresponding author.

**Figure 1**

*Machine-Authored Affective Patterns Evoke Distinct Discrete-Emotion Profiles*

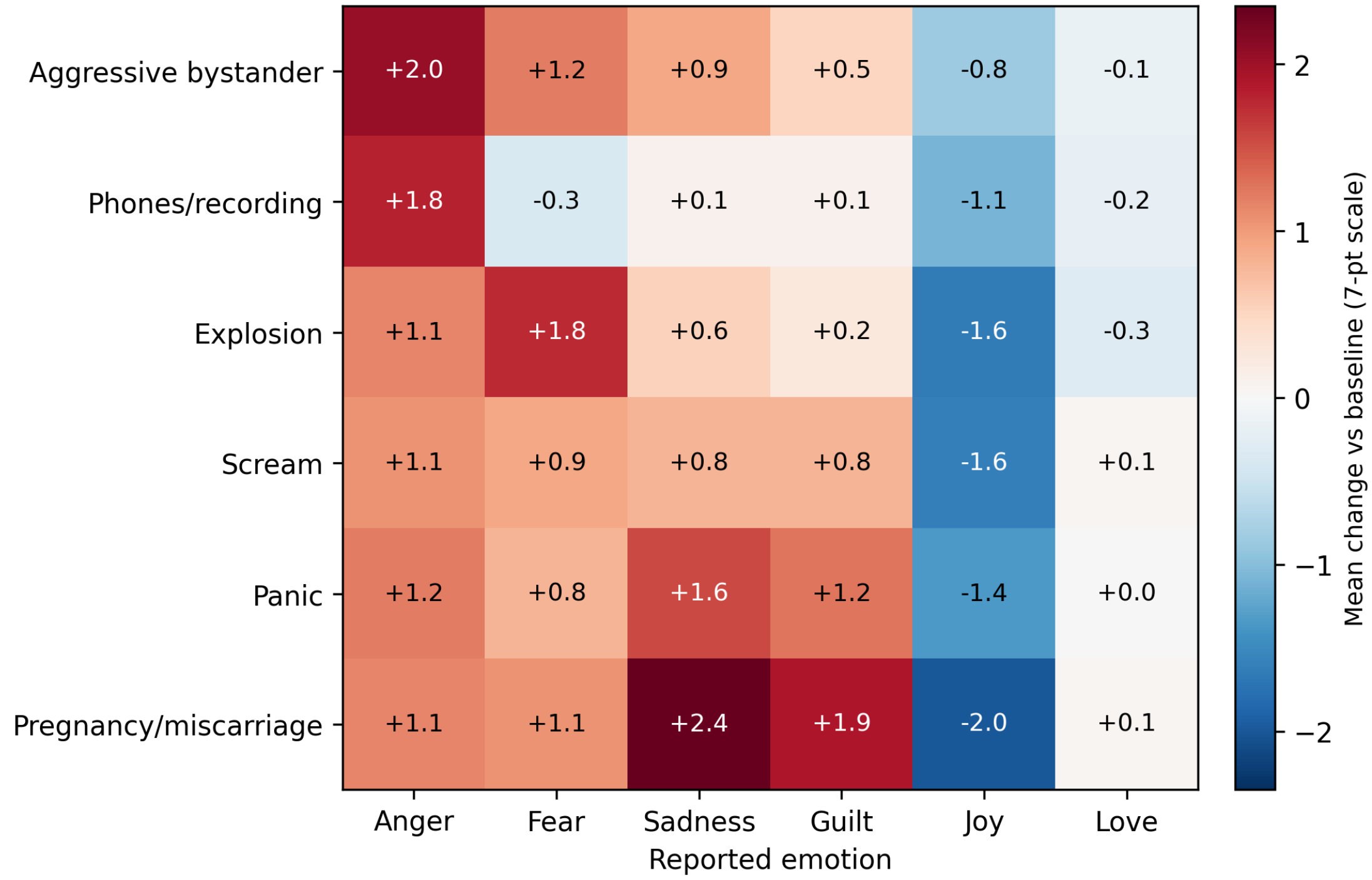

Heatmap of six affective patterns (rows) by six reported emotions (columns), showing mean change from baseline on a 7-point scale. Each pattern has a distinct signature: the aggressive bystander and phones/recording raise anger; the explosion raises fear; panic and the miscarriage raise sadness and guilt; all patterns lower joy. Phones/recording uniquely does not raise fear. The miscarriage shows the largest shifts (sadness +2.4, guilt +1.9, joy −2.0).

## Figure 2

*The Machine Moves Felt Emotion but the Channels It Senses Do Not Track It*

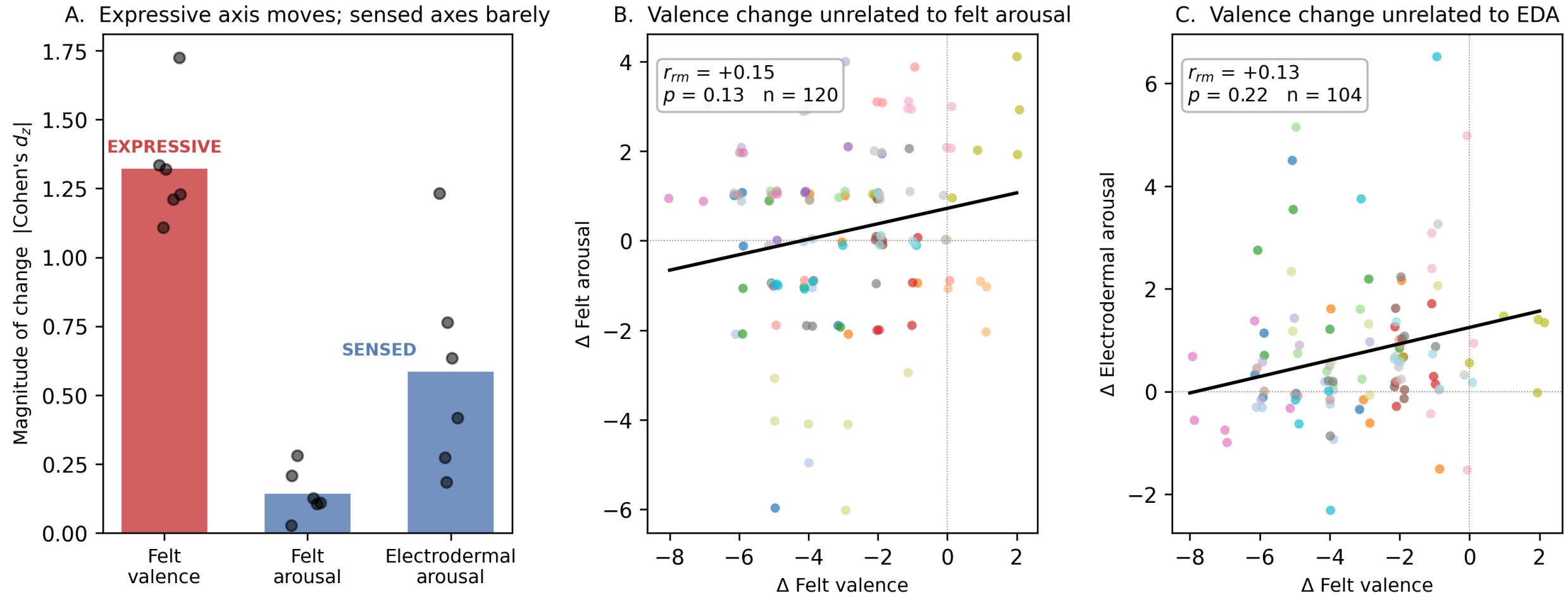

Three-panel figure. Panel A: a bar chart of standardized change magnitude on three channels—felt valence about 1.3, felt arousal about 0.14, and electrodermal arousal about 0.58—showing the expressive channel moves far more than the two sensed channels. Panels B and C: scatterplots of change in felt valence (horizontal axis) against change in felt arousal (B) and electrodermal arousal (C), coloured by participant, each with a near-flat fitted line; the repeated-measures correlations are small and non-significant (r = .15, p = .13; r = .13, p = .22), indicating that the valence change is not tracked by either sensed channel.

**Table 1.** *The expressive channel: change from baseline per pattern (N = 20).*

| Pattern | Valence $d_z$ | Valence $BF_{10}$ | Dominant emotion (Δ) |
|---|---|---|---|
| Aggressive bystander | −1.32 | $2.0 \times 10^3$ | Anger (+2.05) |
| Explosion | −1.34 | $2.3 \times 10^3$ | Fear (+1.75) |
| Panic | −1.11 | $3.1 \times 10^2$ | Sadness (+1.55) |
| Phones/recording | −1.23 | $9.1 \times 10^2$ | Anger (+1.80); fear −0.35 |
| Pregnancy/miscarriage | −1.72 | $5.6 \times 10^4$ | Sadness (+2.35); guilt (+1.90) |
| Scream | −1.21 | $7.8 \times 10^2$ | Anger (+1.05) |

*Note.* All valence contrasts $p < .001$ and survive Holm and Benjamini–Hochberg correction. Discrete-emotion profiles are shown in Figure 1.

**Table 2.** *The sensed channels per pattern.*

| Pattern | Felt arousal $d_z$ ($BF_{01}$) | Electrodermal $d_z$ ($BF_{10}$), $n$ | Mean HR $d_z$ ($BF_{10}$) |
|---|---|---|---|
| Aggressive bystander | +0.11 (3.9) | +0.64 (3.7), 18 | −0.45 (1.2) |
| Explosion | +0.21 (2.9) | **+1.23 (259)**, 17 | **−1.33 (2135)** |
| Panic | +0.28 (2.2) | **+0.76 (9.7)**, 18 | −0.39 (0.8) |
| Phones/recording | +0.03 (4.3) | +0.42 (0.9), 18 | −0.53 (2.1) |
| Pregnancy/miscarriage | −0.13 (3.7) | +0.18 (0.3), 16 | −0.47 (1.3) |
| Scream | +0.11 (3.9) | +0.27 (0.4), 17 | **−0.93 (63)** |

*Note.* Felt arousal is change from the baseline measurement (self-report); electrodermal and heart-rate contrasts are against the per-pattern activity-matched baseline. Boldface marks contrasts surviving Holm correction within the family. $BF_{01} > 1$ indicates evidence *for* no arousal change; $BF_{10} > 3$ indicates moderate evidence *for* an electrodermal or cardiac change.